\documentclass[natbib,10ptpreprint]{elsarticle}
\pdfoutput=1
\usepackage{amsmath,amssymb}
\usepackage{graphicx}
\usepackage[tight]{subfigure}
\usepackage{textcomp}
\usepackage{gensymb}
\usepackage{mathtools}
\usepackage{rotating}
\usepackage{textcomp}
\usepackage{fancyhdr}
\usepackage{hyperref}
\hypersetup{pdfauthor={Gonçalves},pdftitle={Volume dependence of magnetic properties in Co$_2$Cr$_{1-x}Y_x$Ga ($Y$=Ti\--Ni) Heusler alloys: a first-principles study}}
\begin{document}
%%%%%%%%%%%%%%%%%%%%%%%%%%%%%%%%%%TITLE%%%%%%%%%%%%%%%%%%%%%%%%%%%%%%%%%%%%%%%%%
\title{Volume dependence of magnetic properties in Co$_2$Cr$_{1-x}Y_x$Ga ($Y$=Ti\--Ni) Heusler alloys: a first-principles study}
%%%%%%%%%%%%%%%%%%%%%%%%%%%%%%%%%%ABSTRACT%%%%%%%%%%%%%%%%%%%%%%%%%%%%%%%%%%%%%%
\author[ua]{J. N. Gon\c{c}alves}
\ead{joaonsg@ua.pt}
\author[ua]{N. M. Fortunato}
\author[ua,up]{J. S. Amaral}
\author[ua]{V. S. Amaral}
\address[ua]{CICECO - Aveiro Institute of Materials and Departamento de F\'isica, Universidade de Aveiro, 3810-193 Aveiro, Portugal}
\address[up]{IFIMUP and IN-Institute of Nanoscience and Nanotechnology, Rua do Campo Alegre, 678, 4169-007 Porto, Portugal}
	
\begin{abstract}
 
The magnetic properties tuning and volume dependence in the series of quaternary full Heusler alloys with formula Co$_2$Cr$_{1-x}Y_x$Ga ($Y =$ Ti, V, Mn, Fe, Co, Ni) were studied with a detailed first-principles exploration. We employ the density functional KKR method with the coherent potential approximation, estimating effective Heisenberg exchange constants  via the magnetic force theorem together with mean-field Curie temperature ($T_C$) and magnetic moment for compositions in the whole concentration range. The volumetric dependency of these magnetic properties is studied, particularly the pressure derivatives of $T_C$ at equilibrium. Our ternary alloy calculations show good agreement with local-density and generalized gradient approximations in the literature. The quaternary alloys show a wide range of tunable magnetic properties, where magnetic moments range from $0.8$ to $4.9$\,$\mu_B$, $T_C$ from $130$\,K to $1250$\,K, and $dT_C/dV$ values range from $-7$ to $+6.3$\,K\,\AA{}$^{-3}$. 

\end{abstract}

\begin{keyword}
	Heusler alloys\sep Curie temperature\sep pressure\sep volume dependence
\end{keyword}

\maketitle
%%%%%%%%%%%%%%%%%%%%%%%%%%%%%%%%%%%%%%%%%%%%%%%%%%%%%%%%%%%%%%%%%%%%%%%%%%%%%%%%
%%%%%%%%%%%%%%%%%%%%%%%%%%%%%%%%%%%%%%%%%%%%%%%%%%%%%%%%%%%%%%%%%%%%%%%%%%%%%%%%

\section{Introduction}

Heusler alloys are known for the great variety of compositions and properties, since many 
elements can occupy the $X$, $Y$ and $Z$ sites in the L2$_1$ structure with fractional occupations~\cite{graf_simple_2011}, allowing a great flexibility. Improved properties may also be obtained by exploring disorder, non-stoichiometric compositions, and variation with external parameters such as magnetic field, temperature and pressure~\cite{entel_phase_2012}. The exploration of several independent factors can be achieved in a controlled way using first-principles calculations. Interest in Heusler alloys has mainly been concerning electronic structure and transport properties, motivated by application in spintronic devices~\cite{felser_spintronics_2013}, especially in half-Heusler $XYZ$ and full-Heusler Co$_2YZ$ alloys. Other aspects of Heusler alloys have also motivated first-principles studies, including exchange interactions~\cite{jakobsson_first-principles_2015,sasioglu_first-principles_2004}, to calculations of the magnetocaloric effect~\cite{sokolovskiy_achieving_2015,buchelnikov_first-principles_2010} with the aid of Monte Carlo simulations.

In this context, a related aspect which benefits from computation but has 
been less studied is the variation of exchange interactions, and ferromagnetic Curie 
temperature ($T_C$) with volume, or hydrostatic pressure.  An initial detailed theoretical study of 
this subject was concerned with Ni$_2$MnSn~\cite{PhysRevB.71.214412}. More recently, Mn$_2$Co$Z$ ($Z =$ 
Al, Ga, In, Si, Ge, Sn, Sb) compounds have been analyzed~\cite{meinert_exchange_2011}. 
(Ni$_{1-x}T_x$)$_2$MnSn ($T =$ Pd, Cu) alloys have also been studied as a function of concentration 
and pressure~\cite{PhysRevB.84.174422}. For Cu alloying, a crossover behavior was found, with  $dT_C/dP$ 
changing sign as a function of Cu concentration. 

Experimental investigations of $T_C$ with pressure have also been specially focused in the Mn based systems, due to coupling between structure and magnetism (with, e.g.\ the shape memory and magnetocaloric effects), including Ni$_2$Mn$Z$ ($Z =$ Al, Ga, In, Sn, and Sb)~\cite{kaneko_pressure_1981,kanomata_effect_1987}, Au$_2$MnAl and Pd$_2$Mn$Z$ ($Z =$ Sn and Sb)~\cite{Shirakawa1987421}. For all these alloys, wherein Mn has localized moments, the positive $dT_C/dP$ (negative $dT_C/dV$) measured is consistent with an empirical interaction curve as a function of the distance between Mn atoms~\cite{kanomata_effect_1987}.

For cobalt-based (Co$_2YZ$) alloys there have been experiments with Co$_2$TiAl/Co$_2$TiGa~\cite{kanomata_effect_1988,dimasi_pressure_1993}, Co$_2$ZrAl~\cite{Kanomata200526}, and  Co$_2$VAl/Co$_2$VGa~\cite{kanomata_magnetic_2010}. These measurements revealed a negative $dT_C/dP$, which is typical of itinerant magnetism. The Co$_2$Mn$Z$ ($Z =$ Ga, Si, Ge, Sn) alloys, however, display a positive $dT_C/dP$, in conformity with the other Mn-based systems~\cite{kurtulus_electronic_2005}. 
 
In a previous study~\cite{goncalves_magnetovolume_2014} we have made an investigation of different Co$_2YZ$ compounds, including the variation of lattice parameters, and its effect on magnetic interactions and $T_C$. In this work we focus in particular on stoichiometries based in the Co$_2$CrGa compound. Cr is substituted with one of six different elements: Ti, V, Mn, Fe, Co, and Ni, and these alloys are studied with fractional concentrations covering the whole concentration range. 

For this kind of system, previous GGA/LDA calculations show good agreement with experimental data. For example, Ref.~\cite{seema_electronic_2015} uses GGA, Ref.~\cite{PhysRevB.76.024414} is a compilation showing the success of LDA/GGA in Co$_2YZ$ systems, where the moment is well represented as well as $T_C$ (in this case using the random phase approximation). Ref.~\cite{zhang_magnetic_2004} also shows that the LDA spin moment is consistent with experiment. According to the discussion of Ref.~\cite{Thoene}, LDA/GGA is sufficient for accurate Curie temperatures (mean field) and magnetic moments, for several Co$_2YZ$ and two Ni$_2$Mn$Z$ alloys, with the only exception for the case of Co$_2$FeSi (which we do not study here), for which LSDA+U is found to better describe the magnetic moment.

However, with respect to magnetovolume effects, previous calculations have shown significant differences between different exchange-correlation approximations. 
In \textit{bcc}-Fe, $dT_C/dP$ was found to be -1.4 K/GPa with GGA and -6 K/GPa with LDA~\cite{PhysRevB.79.184406}. These results are much smaller than previous first-principles results (16 and 18 K/GPa) with a different method~\cite{PhysRevB.67.012407}, and thus the authors claimed a correct prediction of the experimental $dT_C/dP\approx0$. Nevertheless, these are still significant changes, both relative and absolute. Values of the same order of magnitude have been found for Heusler alloys (for example, 6.2 and 5.6 K/GPa respectively for Ni$_2$MnSn and Ni$_2$MnIn~\cite{austin_effect_1967}). Therefore, we found it necessary to study the influence of the exchange-correlation approximation in a series of LDA and GGA calculations.  We will show that, for the compositions containing Ti and V, $dT_C/dV$ is strongly dependent on the exchange-correlation functional: the measured sign is obtained with the local density approximation, but not with the generalized gradient approximation. 

Using the LDA, we find that those compositions close to Co$_2$TiGa and Co$_2$VGa display positive (negative) $dT_C/dV$ ($dT_C/dP$), as measured, and expected for itinerant magnetism. Moving forward in the transition metal period to Cr, Mn, and Fe, $dT_C/dV$ ($dT_C/dP$) becomes negative (positive), consistent with localized moments.

To our knowledge Co$_2$NiGa forms in a disordered \textit{fcc} or ordered Pt$_2$FeCu-type structure~\cite{dai_structure_2007}, while Co$_3$Ga is not known in any form. We will also analyze the alloying of Co and Ni in the quaternary alloys with the L2$_1$ structure, bearing in mind that we are not describing known compounds in these cases. 
A previous first-principles study has already analyzed the stability of Co$_2$NiGa, and found, as expected, that the ground state cubic structure is metastable with respect to the tetragonal distortion~\cite{arroyave_investigation_2010}. However, L2$_1$-Co$_2$NiGa may be formed with novel synthesis methods, and the fractional alloys with Cr should be synthesized more easily than the ternary compounds, especially for low substitution concentrations. For the compositions close to Co$_3$Ga and Co$_2$NiGa our calculations show  $dT_C/dV$ ($dT_C/dP$) again positive (negative). Therefore, substituting Ti, V, Co, or Ni for Cr gradually decreases $dT_C/dV$ and eventually changes its sign.

\section{Calculation Details}
We use the KKR Green's function density functional theory, as implemented in the SPR-KKR code~\cite{Ebert,Ebert2}. The full potential\footnote{The use of the full potential by Thoene et al.~\cite{Thoene} was found important in some cases to obtain accurate Curie temperatures.} spin-polarized scalar-relativistic method is used.  As the exchange-correlation approximation we consider the generalized gradient approximation (GGA) with the Perdew-Burke-Ernzerhof parameterization~\cite{Perdew97} and the local density approximation (LDA) with the Vosko-Wilk-Nusair parameterization~\cite{VWN} for the ternary compounds. The LDA is used for the detailed study of the alloys. The reciprocal space is sampled with a $22\times22\times22$ $\mathbf{k}$-mesh, the calculation in the complex energy path uses 32 points, and the angular momentum cutoff is $l=3$. The magnetic order is assumed ferromagnetic, which is consistent with the known measured cases, such as the ternary compounds (also under pressure), and with Fe or Mn doping~\cite{Umetsu2005,umetsu_magnetic_2015}. 
For the quaternary, fractional alloys, we sample variation of different concentrations of transition metal element substitutions in the $Y$ site of the L2$_1$ structure, with chemical formula Co$_2$Cr$_{1-x}Y_{x}$Ga, and considering six neighbor elements ($Y$=Ti, V, Mn, Fe, Co, Ni). The coherent potential approximation is used to simulate the disorder in that site~\cite{soven_coherent-potential_1967}.~\footnote{It was previously shown~\cite{ozdogan_engineering_2008}, by comparison with supercell calculations, that short-range interactions should not affect the magnetic properties significantly for the type of alloys we are considering.} Steps of $\Delta x=0.1$ were considered. 

Self-consistent total energy calculations at different volumes are used with the Murnaghan equation of state~\cite{Murnaghan} to fit the energy-volume curve and find the corresponding pressures. (There are no internal relaxations involved due to the symmetries of the L2$_1$ structure.)

The exchange constants of an effective 
Heisenberg model are calculated with the magnetic force theorem of Liechtenstein et al.~\cite{mft}. For this calculation a cluster of radius four times larger than the $L2_1$ cell lattice constant is considered around each site and all the pairwise interactions in that cluster are calculated, considering the four sublattices.  
$T_C$ is determined in the mean field approximation for a multisublattice system~\cite{Anderson,PhysRevB.71.214412}, i.e., using the highest eigenvalue of the matrix $\mathbf{J}$ containing the pairwise exchange interaction sums.

\section{Results}

\subsection{Ternary compounds: exchange-correlation approximation and comparison with experiments}

The calculated lattice parameters and  magnetic moments are shown and compared with experiments~\cite{kanomata_magnetic_2010,zhang_magnetic_2004,Umetsu2005,sasaki_magnetic_2001,webster_magnetic_1971} in Table~\ref{tab:TCLDAGGA}. GGA is more accurate for the lattice parameters. The bulk moduli are also shown, and follow the usual tendency of larger values for LDA, with smaller lattice parameters. With regards to the magnetic moments, there is not one approximation consistently better for all compounds: LDA is better for Co$_2$TiGa, GGA is better for Co$_2$FeGa and slightly better for Co$_2$CrGa, while for the other compounds both approximations have similar accuracies. For all compositions except Co$_2$CrGa, the $T_C$ obtained with LDA is closer to the experimental~\cite{sasaki_magnetic_2001,buschow_magnetic_1981,umetsu_magnetic_2004,umetsu_magnetic_2008,Umetsu2005} one than that obtained with GGA. This does not necessarily mean that LDA produces a better estimate of the exchange constant values than GGA, since the mean field approach used here usually  overestimates $T_C$.

Although the mean field approach to Curie temperatures usually overestimates Curie temperatures, Thoene et al.~\cite{Thoene} have obtained $T_C$ close to the experimental values for a variety of Heusler compounds. For the case of Co$_2$CrGa, it was even underestimated, 366\,K with respect to the experimental value of 495\,K~\cite{umetsu_magnetic_2004}. However, this comes from using the experimental lattice parameters instead of the theoretical ones. As shown in Tab.~\ref{tab:TCLDAGGA}, the $T_C$ calculated at the theoretical LDA lattice is close to the experimental one. We also find similar values of the derivatives $dT_C/dV$ with both approximations at the equilibrium values.

\begin{sidewaystable}
	\caption{Comparison of the calculated (GGA and LDA) equilibrium lattice parameters $a_0$ (\AA{}), bulk moduli $B$ (GPa),  Curie temperature $T_C$ (K),  magnetic moment $m\,$($\mu_B$), $dm/dV$ ($10^{-3}\mu_B$\,\AA{}$^{-3}$), and $dT_C/dV$ (K\,\AA{}$^{-3}$) at the equilibrium lattice parameters, and experimental $T_C$ and $a_0$.}
	\begin{tabular}{l|c|c|c|c|c|c|c|c|c|c|c|c|c|c|c}
		       & \multicolumn{6}{c|}{GGA}            & \multicolumn{6}{c|}{LDA}        & \multicolumn{3}{c}{exp.}    \\ \hline
		Compound   & $a_0$  & $B$  &$T_C$ & $m$ & $dT_C/dV$  &$dm/dV$ &$a_0$&B&$T_C$  & $m$ &$dT_C/dV$ &$dm/dV$ & $a_0$ &$T_C$ & $m$      \\ \hline\hline
		Co$_2$TiGa & 5.87 & 179 &  243 &   0.97 &-1.2 &-0.2 & 5.72   & 216 &   130 & 0.82 & 4.5   & 5.1 &5.86  & 128    &0.82\\ \hline
		Co$_2$VGa  & 5.80 & 194 &  469 &   1.96 &-1.5 &-2.8& 5.66   & 237 &   345 & 1.96 & 1.7   &-1.2& 5.78 & 352 & 2.04  \\ \hline
    	Co$_2$CrGa & 5.75 & 195 &  485  & 2.99 &  -5.3 &9.9& 5.61  & 247 &   470 & 2.96 &-4.6   &0.4& 5.81 & 495  & 3.01  \\ \hline
		Co$_2$MnGa & 5.75 & 187 &  662 &   4.09 &-8.2 &13.8 & 5.60   & 237 &   670 & 4.01 &-7.0&9.2   & 5.77  & 685 & 4.05\\ \hline
		Co$_2$FeGa & 5.75 & 187 & 1317 &   4.99 &-4.9 &17.4& 5.60   & 231 &   1252& 4.87 &-3.2&26.5   & 5.74 & 1093  &  5.15\\ \hline
	\end{tabular}
	\label{tab:TCLDAGGA}
\end{sidewaystable}

For other compositions, $dT_C/dV$ does not always show such a good agreement between functionals. This is shown here for Co$_2$TiGa, Co$_2$VGa, Co$_2$MnGa, and Co$_2$FeGa. For Co$_2$MnGa and Co$_2$FeGa, similarly to Co$_2$CrGa, there are quantitative differences but the results are still consistent, since the derivatives are not far in the two cases and have the same sign. For Co$_2$VGa, however, the $T_C$ dependence changes sign with the different approximations, at the theoretical lattice parameters. 
 The difference of the derivative between approximations is not far from the other cases, but some factors contribute to the change of sign: small values near the theoretical parameters, and the large variation of the value with volume in LDA, plus the usual overbinding of LDA. 
For Co$_2$TiGa, the volume dependence of the Curie temperature is even more distinct for the two functionals, an almost constant behavior with negative slope with GGA, but a steeper, positive dependence with LDA.  (Note: $dT_C/dV$ values were calculated with respect to the L2$_1$ unit cell volume differences. If we used the volume per formula unit, the values would be four times larger.)

It is also interesting to analyze the change of the magnetic moments with volume. For $Y$=Mn and Fe the results are consistent for both approximations only with small quantitative differences, as in $dT_C/dV$, but with opposite sign. For the case of Co$_2$TiGa we have exactly the same sign change as in $dT_C/dV$, indicating the change of $m$ as the main factor in the variation of the exchange interactions. In contrast, for Co$_2$VGa $dm/dV$ has the same sign between functionals, opposing the $dT_C/dV$ behavior. For Co$_2$CrGa, $dm/dV$ is much higher for GGA than for LDA (where it is almost zero since $m$ is a minimum close to equilibrium). This also contrasts with the $T_C$ dependence, which is basically the same, negative and not insignificant, for both functionals.

The $T_C$ values and its variation with volume and approximation (LDA/GGA) can be explained as a result of the underlying exchange interactions. The two largest exchange interactions are shown in Table~\ref{tab:exchg}. For Co$_2$FeGa and Co$_2$MnGa, $J_{Co-Y}$ is much larger than the next largest interaction, Co-Co, and both interactions decrease with volume, consistent with $T_C$. $J_{Co-Fe}$ decreases slightly faster for GGA, as with $T_C$. For Co$_2$CrGa, $J_{Co-Cr}$ is only slightly larger than the Co-Co interaction. The dependence of the first interaction is slightly steeper for LDA, consistent with the small quantitative difference between approximations. Co$_2$VGa and Co$_2$TiGa are different, since here the first Co-Co is much larger than the first Co-$Y$ interaction, which is already insignificant. For Co$_2$VGa the first Co-Co interaction has the same variation in both approximations, although with GGA the variation is less steep, and it is not enough to account for the sign change with d$T_C$/d$V$ (it must be accounted by considering Co-Co at larger distances, which also have distinct volume changes between LDA and GGA). For Co$_2$TiGa the first two Co-Co interactions show the same behavior as $T_C$. The distinct behaviors between functionals for Y=Ti, V can then be ascribed to the different descriptions of Co-Co interactions.

\begin{table}
	\caption{Two largest pairwise exchange interactions (meV) for the ternary compounds Co$_2Y$Ga ($Y$=Ti, V, Cr, Mn, Fe), calculated with LDA and GGA (in parenthesis). $J^1$ is $J_{Co-Y}$ for $Y$=Cr, Mn, Fe, and $J_{Co-Co}$ for $Y$=Ti, V.\label{tab:exchg}}
	\begin{tabular}{l|c|c}
		Compound      &     $J^1$   & $J^2_{Co-Co}$         \\ \hline\hline
		Co$_2$TiGa    & 2.56 (4.59)           & 1.26 (2.35)                     \\ \hline 
		Co$_2$VGa     & 6.74 (8.29) & 1.54 (2.17)           \\ \hline
		Co$_2$CrGa    & 4.55 (5.88) & 3.72 (4.35)           \\ \hline
		Co$_2$MnGa    & 10.62 (11.50) & 1.49 (1.92)         \\ \hline
		Co$_2$FeGa    & 21.75 (22.62)           & 2.30 (2.56)             \\ \hline
	\end{tabular}
\end{table}

Table~\ref{tab:expthe} summarizes present experimental knowledge of $dT_C/dP$ in our alloys, including Co$_2$TiGa and Co$_2$VGa, along with our results. 
The sign is correct, but only if we consider the LDA calculations, as the GGA calculations give the opposite sign. 
Even considering the LDA results, the calculated values are $\sim3$ and $\sim5$ times smaller for Co$_2$TiGa and Co$_2$VGa, respectively. 

The main reason for this disagreement is unclear at present. Alling et al.\ have suggested that to improve $T_C$ estimates it should be necessary to consider also finite-temperature effects, such as magnetic excitations and chemical disorder~\cite{alling_effect_2009}. In that work, it was also shown that the use of the disordered local moment state as the reference for the calculation of exchange interactions may bring some change in the dependence of $T_C$ with volume. For Ni$_2$MnSn~\cite{PhysRevB.71.214412}, the calculated value 3.22 was also smaller than the experimental 7.44 (K$/$GPa), and it was calculated that intersublattice interchange between Mn and Ni can strongly increase $dT_C/dP$. A similar antisite disorder may affect the samples discussed here. Theoretical improvements may also play a role, such as more sophisticated Curie temperature calculation approaches (such as the random phase approximation) and, as we have shown here, different exchange-correlation functionals.

\begin{table}
\caption{Comparison of the measured and calculated (LDA and GGA) $dT_C/dP$ (K$/$GPa) in Co$_2$VGa and Co$_2$TiGa. \label{tab:expthe}}
\begin{tabular}{l|c|c|c}
Compound     & Measured                                                                  & LDA     & GGA\\ \hline\hline
Co$_2$TiGa   & $-13$~\cite{kanomata_effect_1988}; $-12.7$~\cite{sasaki_magnetic_2001}    & $-4.0$  & $+0.4$  \\ \hline % +1-3 GGA?
Co$_2$VGa    & $-7.8$~\cite{kanomata_magnetic_2010}                                          & $-1.5$  & $+1.2$   \\ \hline
%Co$_2$TiAl   & $+6$~\cite{sasaki_magnetic_2001}   ; $-7\pm 2$~\cite{dimasi_pressure_1993}& $-7.0$  & $+0.8$   \\ \hline
\end{tabular}
\end{table}

We will continue with a detailed study of the fractional alloys using LDA, due to its better agreement with experiment for $dT_C/dP$.

\subsection{Quaternary alloys}

\begin{figure}%
  \centering 
\subfigure{\includegraphics[width=0.80\linewidth]{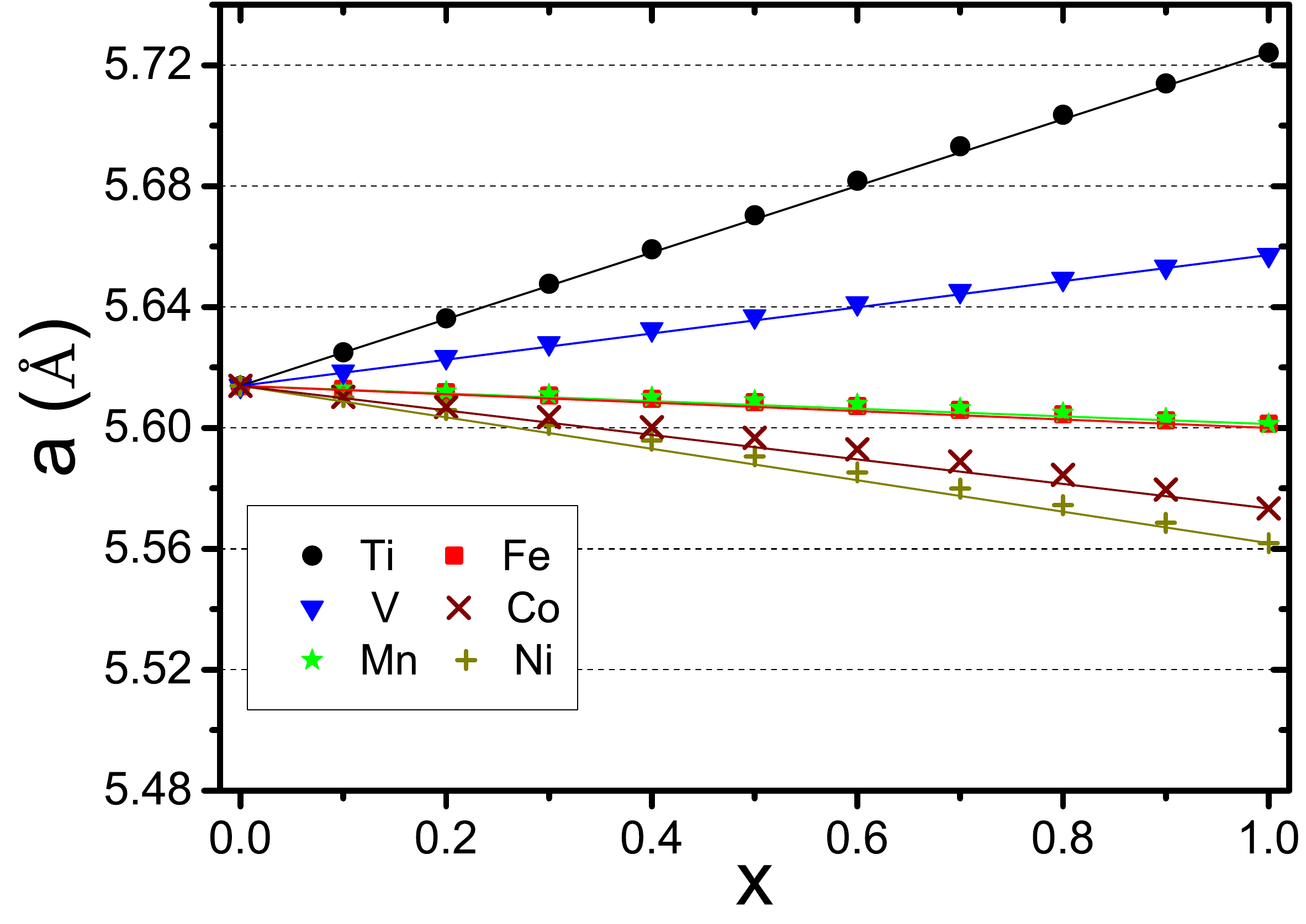}}
\caption{Theoretical lattice parameters for different compositions Co$_2$Cr$_{1-x}Y_x$Ga ($Y$=Ti, V, Mn, Fe, Co, Ni). The lines connect the endpoints. (The Mn and Fe points are superposed in the graph, showing almost the same values.)} 
\label{fig:latt}%
\end{figure}

Fig.~\ref{fig:latt} shows the optimized lattice parameters for all the compositions studied. 
The variation of the lattice parameter with the concentration of the transition metal is approximately linear.

In what follows we will consider the magnetic properties, in order of atomic number for the six neighbor elements of Cr which we considered in our calculations.

\begin{figure}
	\subfigure[]{
		\includegraphics[width=0.45\linewidth]{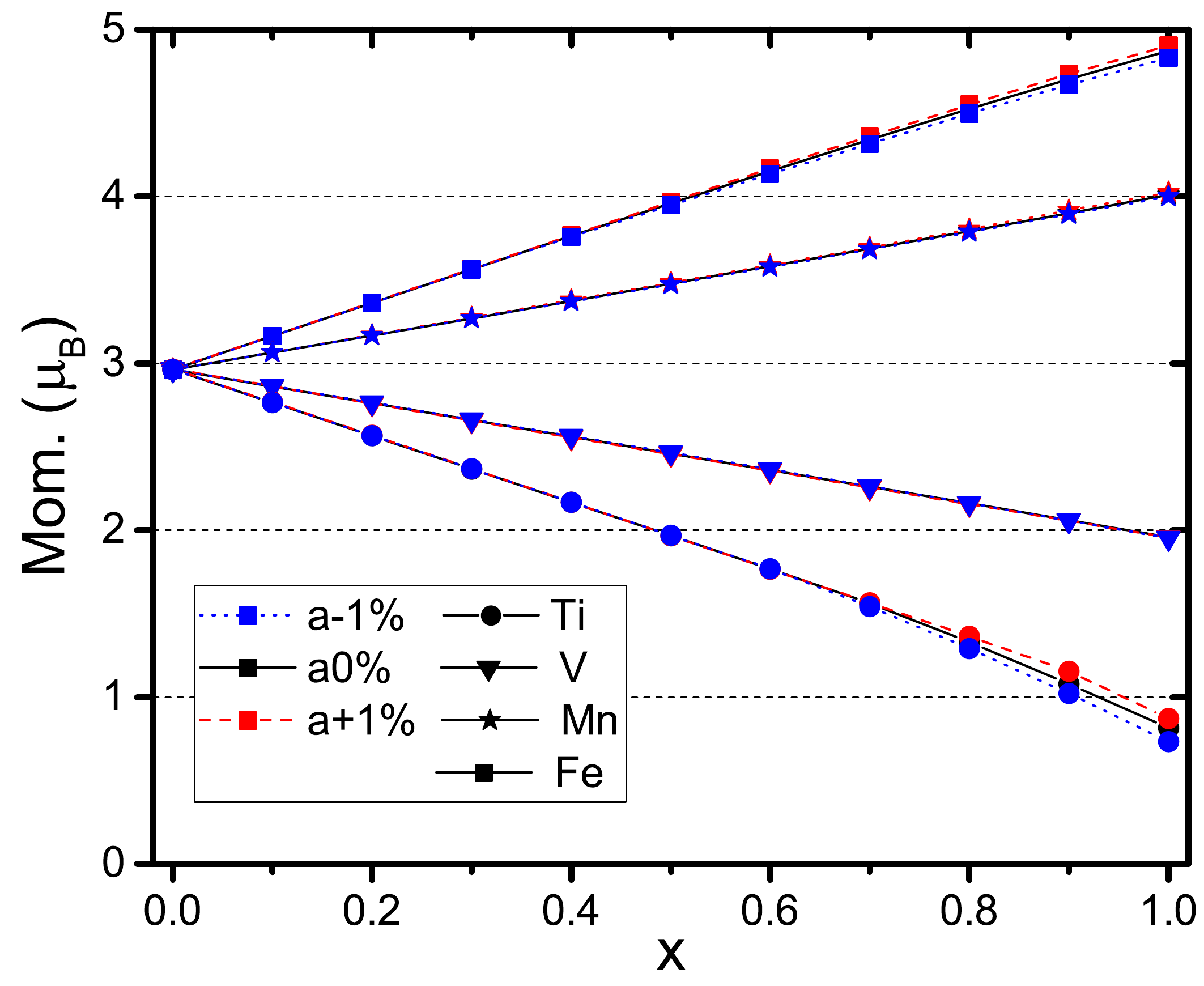}\label{fig:spin_1}
	}
	\subfigure[]{
		\includegraphics[width=0.45\linewidth]{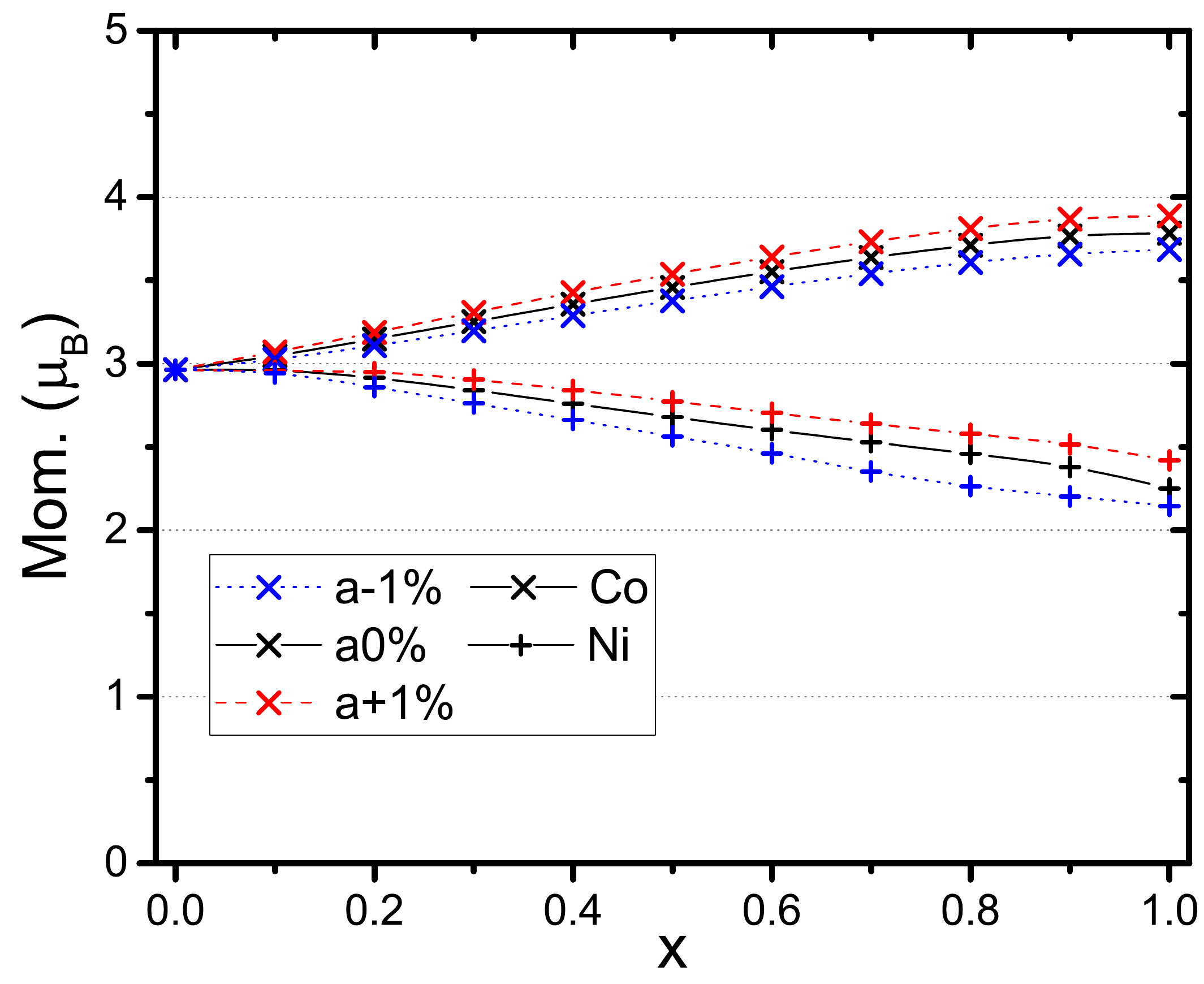}\label{fig:spin_2}
	}
	\centering
	\caption{Magnetic spin moments ($\mu_B$) for the alloys Co$_{2}$Cr$_{1-x}Y_x$Ga ($Y$ = Ti, V, Mn, Fe, Co Ni), as a function of the concentration $x$ for different calculated lattice parameters around the equilibrium values.} 
	\label{fig:spin}
\end{figure}

Figs.~\ref{fig:spin_1} and ~\ref{fig:spin_2} show the magnetic spin moments for small volume variations (lattice parameter with $\pm1\%$) around the theoretical equilibrium values. The approximate linear variations as a function of the valence electron concentration are in line with a generalized Slater-Pauling rule (also observed in many other Co-based full-Heusler compounds~\cite{galanakis_slater-pauling_2002}) with the slight deviations for the Ti and Fe rich compositions showing respectively slopes $dM/dn_{el}>1$ and $dM/dn_{el}<1$, resulting in moments slightly less than $1\,\mu_B$ and $5\,\mu_B$ respectively.
 Alloying with Ti strongly reduces the spin moment. Co$_2$TiGa has a moment close to $0.8\,\mu_B$ at the theoretical lattice parameter, and we have further calculated that with smaller lattice parameters the magnetic moment falls quickly (not shown) and at a critical pressure of about 30 GPa it vanishes.

For compounds close to the Cr and V rich compositions the magnetic moment is insensitive (changing less than $0.01\,\mu_B$) to the volume changes considered ($\pm1\%$ in the lattice parameter). It is also almost insensitive (changing $0.02\,\mu_B$) for Mn rich compositions. For Fe rich compositions it becomes slightly sensitive ($0.07\,\mu_B$). 
For high concentrations of Ti and even at low concentrations of the other transition metals (Co, Ni) the spin moments become significantly (more than $0.1\,\mu_B$) dependent on volume. The spin moment with full Co substitution is close to $4\,\mu_B$, consistent with the decrease of moment with electrons after Fe ($\approx5\mu_B$), following the rule $M=34-Z$ for itinerant metals~\cite{fecher_slater-pauling_2006}. Co$_2$NiGa follows the same decreasing trend, but reaches a lower value than expected, closer to $2 \mu_B$ than $3\,\mu_B$.

In all cases, at a given composition the dependence is such that higher volumes always lead to higher moments. The variation of spin moments with lattice parameters is usually negligible when compared with the changes that can be achieved with changing composition. However, this is not so in the compositions close to Co$_2$TiGa, where a relatively small pressure can decrease the moments significantly, and for compositions close to the Co$_3$Ga and Co$_2$NiGa cases, where a change in the lattice parameter of $1\%$ can produce moment changes of $0.15\,\mu_B$, comparable to a change of $x=0.3$ in composition.

\begin{figure*}
	\centering
	\subfigure[]{
		\includegraphics[width=0.45\linewidth]{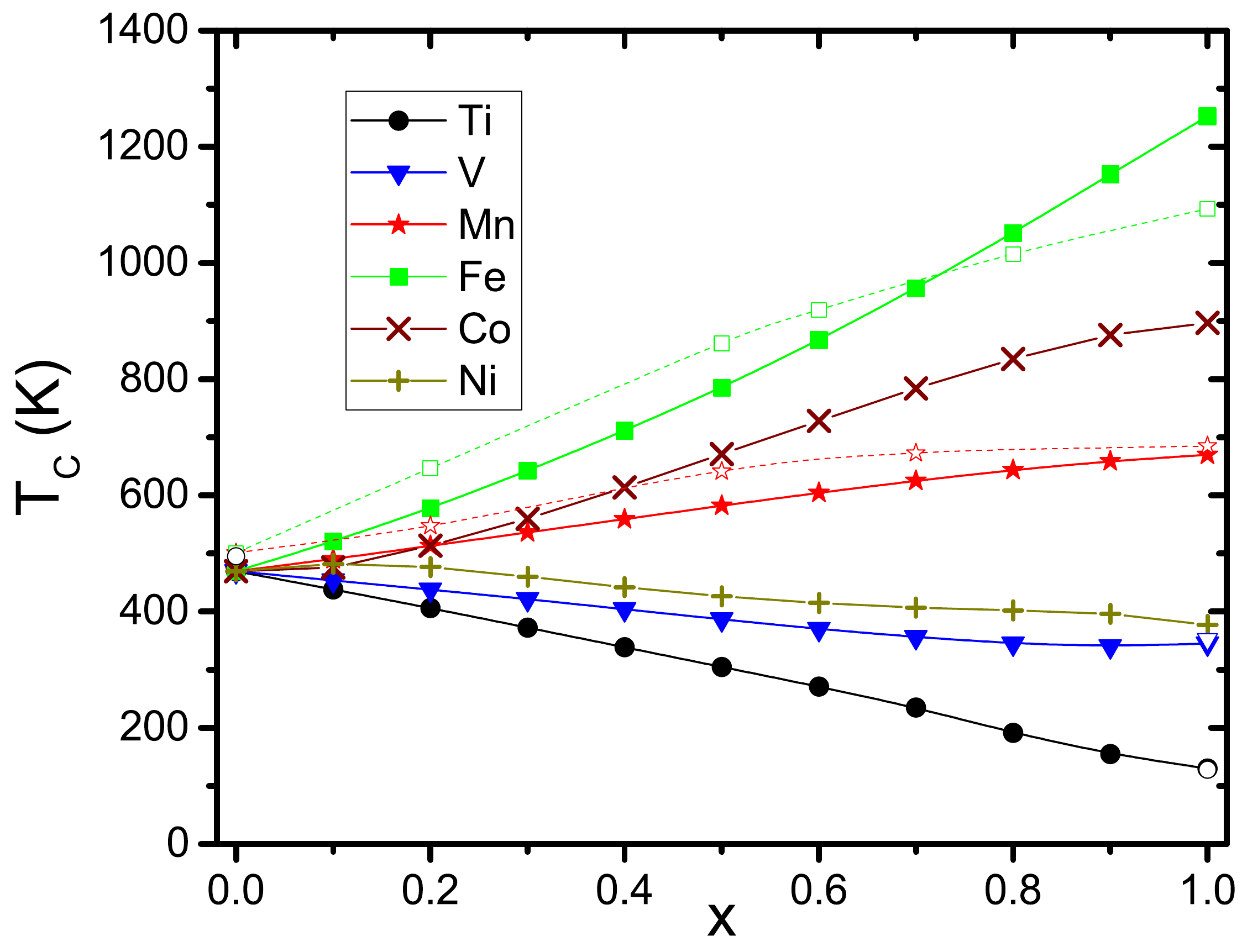}\label{fig:TC_aeq}
	}
	\subfigure[]{
		\includegraphics[width=0.45\linewidth]{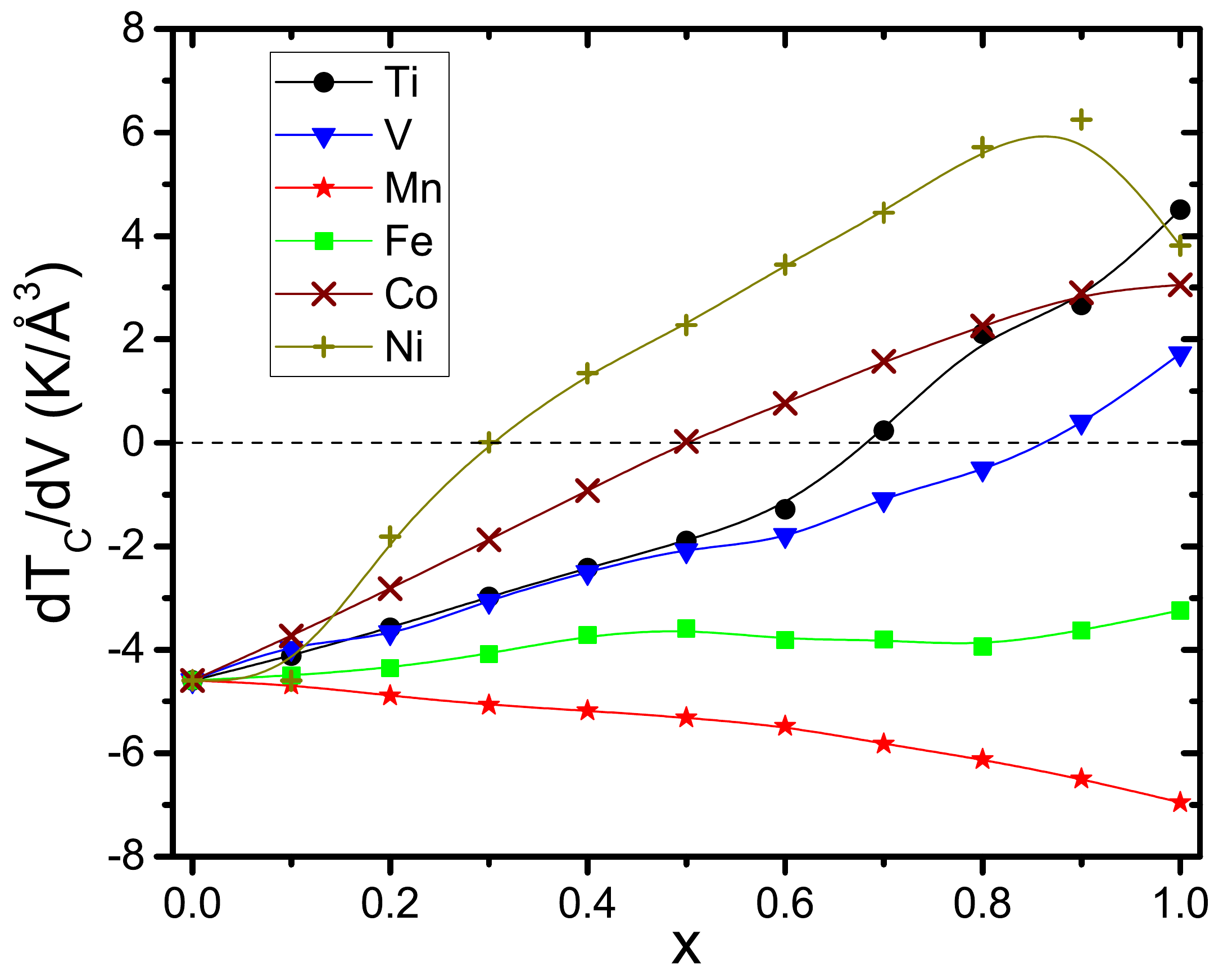}\label{fig:dTCdV:aeq}
	}
	\caption{a) Curie temperature (K) for the alloys Co$_{2}$Cr$_{1-x}Y_x$Ga ($Y$=Ti, V, Mn, Fe, Co, Ni) at the equilibrium lattice parameters, as a function of the element concentration. The open symbols correspond to experimental values.  
		b) corresponding derivative of the Curie temperature with respect to volume (K\,\AA{}$^{-3}$). The lines are guides to the eye.}
	\label{fig:TCs:aeq}
\end{figure*}

The calculated Curie temperatures at the equilibrium lattice parameters %and $dT_C/dV$ 
 are shown in Fig.~\ref{fig:TC_aeq}. 
$dT_C/dV$ is shown in Fig.~\ref{fig:dTCdV:aeq}, also for the equilibrium lattice parameters. For the Ti case it shows negative values on the Cr rich side  and positive values on the Ti rich side. At $x_{\text{Ti}}\approx0.6$ (where $T_C$ is close to 300\,K), $dT_C/dV$ is very small. 
 With V substitution $T_C$ also falls, and similarly $dT_C/dV$ also changes sign, in this case at V rich compositions ($dT_C/dV\approx0$ at $x\approx0.8$). With V doping, it is possible to get $T_C$ close to room temperature ($\sim 350$\,K) at high vanadium concentrations,   
and with compression 
it is possible to further reduce $T_C$. With Mn doping %the Curie temperature increases regularly by 120\,K, with 
$dT_C/dV$ becomes more negative. In spite of the great range of Curie temperatures for Fe alloying, the respective change with volume is the most constant in all the substitutions. With Co, $T_C$ increases from 450 to 900\,K. Just as with Ti and V, $dT_C/dV$ changes sign from negative to positive, and at $x_{\text{Co}}\approx0.5$, $dT_C/dV\approx0$. With Ni  $T_C$ changes little, 
 while $dT_C/dV$ changes from negative to positive with increasing Ni concentration, reaching the highest values for Ni-rich compositions. 

$T_C$ varies monotonically with concentration $x$ of the elements, with no anomalies between the extremes, except where the minimum $T_C$ is achieved slightly away from full V substitution, and a local $T_C$ maximum for low Ni concentrations. $dT_C/dV$ usually also presents extremes at the endpoints, except for Fe and Ni substitutions. In the case of Fe the derivative is not monotonic, nevertheless varying little, in the interval [-4.4,-3.2]\,K\,\AA{}$^{-3}$. For Ni the maximum and minimum derivatives are reached slightly away from the endpoints. 
  
For the cases of Fe and Mn doping there are previous experimental measurements and calculations, showing a progressive and large  increase of the Curie temperature~\cite{Umetsu2005,umetsu_magnetic_2015}, consistent with our results, although there are differences in the values and curvature of the $T_C(x)$ dependence. For the end compounds there are also measurements of $T_C$~\cite{Umetsu2005,sasaki_magnetic_2001,webster_magnetic_1971,umetsu_magnetic_2008}, and we find very good agreement with our results, except for Co$_2$FeGa, where the calculated value is higher.

The suppression of magnetism with pressure in the (Ti,V) compounds is concomitant with a high positive $dT_C/dV$, with the largest value achieved for Co$_2$TiGa, which is nevertheless smaller in magnitude than the opposite sign value $-7$ achieved in Co$_2$MnGa. However, the (Ti,V) compounds present values of $T_C$ closer to room temperature, which may be useful for devices, with applications in room temperature refrigeration or spintronics, for example. Finally, we also predict that Co$_2$Cr$_{1-x}$Ni$_x$Ga alloys present high values of $dT_C/dV\approx6$ near the rich Ni side, if these composition are synthesized in the L2$_1$ structure, while Co rich compositions reach values close to Co$_2$TiGa.

\section{Conclusions}

We studied magnetic and magnetovolume coupling properties of fractional Co$_2$Cr$_{1-x}Y_x$Ga alloys ($Y$=Ti, V, Mn, Fe, Co, and Ni). The equilibrium lattice parameters follow approximate linear variations with changing element concentrations, as well as the magnetic moment (in line with the generalized Slater-Pauling rule, due to the addition/subtraction of valence electrons) except for $Y$=Co and Ni. Ti, V, Co and Ni substitutions change the sign of $dT_C/dV$ with respect to the parent Co$_2$CrGa compound. However, at least for the Ti and V cases the values are dependent of the exchange-correlation approximation, and the measured sign is only found with the LDA, which is related to the different descriptions of Co-Co exchange interactions.

The wide $T_C$ range of the studied compositions ($100\--1300$\,K) and range of magnetovolume couplings ($dT_C/dV$ from $-7$ to $+6$\,V\,\AA{}$^{-3}$) establishes the rich behavior of this compound family. Optimization with other elements or different substitutions, disorder and volume optimization may be performed to find alloys with optimized $T_C$ and $dT_C/dV$ values for specific applications (for example, the magnetocaloric effect benefits from a high $|dT_C/dV|$~\cite{rps}).  Further experimental work would be valuable to compare with theory and study the effects the various approximations on $dT_C/dV$, which was shown to be very sensitive to the theoretical approach employed.

\section*{Acknowledgments}

This work was supported by the project RECI/CTM-CER/0336/2012 co-financed by FEDER, QREN reference COMPETE: FCOMP-01-0124-FEDER-027465 and was developed within the scope of the project CICECO-Aveiro Institute of Materials, POCI-01-0145-FEDER-007679 (FCT Ref. UID/CTM/50011/2013), financed by national funds through the FCT/MEC and co-financed by FEDER under the PT2020 Partnership Agreement. We acknowledge FCT grants SFRH/BPD/111270/2015 (J.\ S.\ Amaral) and SFRH/BPD/82059/2011 (J.\ N.\ Gon\c{c}alves).

\bibliography{heusler_2} 
\bibliographystyle{elsarticle-num}

\end{document}